\documentclass[10pt,conference]{IEEEtran}
\IEEEoverridecommandlockouts
\usepackage[]{collab}
\usepackage{cite}
\usepackage{amsmath,amssymb,amsfonts}
\usepackage{algorithmic}
\usepackage{graphicx}
\usepackage{textcomp}
\usepackage{xcolor}

\usepackage{CJKutf8}
\usepackage{makecell}
\usepackage{subcaption}
\usepackage{xspace}
\usepackage{float}
\usepackage{tipa}

\usepackage{multirow}
\usepackage{multicol}
\usepackage{enumitem}
\usepackage{url}
\usepackage{colortbl}
\usepackage{tcolorbox}
\usepackage{booktabs}

\usepackage[T1]{fontenc}

\collabAuthor{jz}{magenta}{Jiazhen Gu}

\newcommand{\ie}{{\em i.e.},\xspace}

\def\BibTeX{{\rm B\kern-.05em{\sc i\kern-.025em b}\kern-.08em
    T\kern-.1667em\lower.7ex\hbox{E}\kern-.125emX}}
\begin{document}

\newcommand{\methodname}{OASIS\xspace}

\title{An Image is Worth a Thousand Toxic Words: \\ A Metamorphic Testing Framework for Content Moderation Software}

\author{
\IEEEauthorblockN{Wenxuan Wang\IEEEauthorrefmark{1}, Jingyuan Huang\IEEEauthorrefmark{1}, Jen-tse Huang\IEEEauthorrefmark{1}, Chang Chen\IEEEauthorrefmark{1}, Jiazhen Gu\IEEEauthorrefmark{1}\thanks{Jiazhen Gu is the corresponding author.}, Pinjia He\IEEEauthorrefmark{2}, and Michael R. Lyu\IEEEauthorrefmark{1}}
\IEEEauthorblockA{
    \IEEEauthorrefmark{1}
    Department of Computer Science and Engineering, The Chinese University of Hong Kong, Hong Kong, China\\
    \IEEEauthorrefmark{2}
    School of Data Science, The Chinese University of Hong Kong, Shenzhen, Shenzhen, China\\
}
}

\maketitle

\begin{abstract}
The exponential growth of social media platforms has brought about a revolution in communication and content dissemination in human society.
Nevertheless, these platforms are being increasingly misused to spread toxic content, including hate speech, malicious advertising, and pornography, leading to severe negative consequences such as harm to teenagers' mental health.
Despite tremendous efforts in developing and deploying textual and image content moderation methods, malicious users can evade moderation by embedding texts into images, such as screenshots of the text, usually with some interference.
We find that modern content moderation software's performance against such malicious inputs remains underexplored.
In this work, we propose \textit{\methodname}, a metamorphic testing framework for content moderation software.
\methodname employs 21 transform rules summarized from our pilot study on 5,000 real-world toxic contents collected from 4 popular social media applications, including Twitter, Instagram, Sina Weibo, and Baidu Tieba.
Given toxic textual contents, \methodname can generate image test cases, which preserve the toxicity yet are likely to bypass moderation.
In the evaluation, we employ \methodname to test five commercial textual content moderation software from famous companies (\ie Google Cloud, Microsoft Azure, Baidu Cloud, Alibaba Cloud and Tencent Cloud), as well as a state-of-the-art moderation research model.
The results show that \methodname achieves up to 100\% error finding rates.
Moreover, through retraining the models with the test cases generated by \methodname, the robustness of the moderation model can be improved without performance degradation.
\end{abstract}


\section{Introduction}
\label{sec-introduction}

In the last decade, there has been a significant proliferation of social media platforms and community forums, leading to a remarkable advancement in contemporary textual communication and content dissemination on a global scale.
Facebook is reaching 3 billion monthly active users in 2023, while the number of Instagram is 2 billion~\cite{mau-social-media}.
However, these platforms inevitably provide malicious users an avenue to spread toxic content due to the anonymity of the web.
In general, toxic contents can be roughly categorized into three major kinds of information~\cite{Wang2023MTTMMT}:
(1) \textit{Abusive language and hate speech}, which are abusive contents targeting specific individuals, such as politicians, celebrities, religions, nations, and the LGBTIQA+~\cite{Badjatiya2017DeepLF};
(2) \textit{Malicious advertisement}, which are online advertisements with illegal purposes, such as phishing and scam links, malware download, and illegal information dissemination~\cite{Li2012KnowingYE}; and
(3) \textit{Pornography}, which is often sexually explicit, associative, and aroused~\cite{Rowley2006LargeSI}.

The presence of toxic contents can cause severe adverse effects.
\textit{Hate speech} exposure among children and adolescents poses a higher chance of victimization and perpetration~\cite{kansok2022systematic}.
\textit{Malicious advertisements} continue to be a significant global problem, resulting in 3.4 billion phishing emails daily and an average of \$4.91 million in breach costs per year~\cite{phishing2023}.
\textit{Pornography} exposure, particularly among younger children, may be disturbing or upsetting and cause significant undesirable effects on the physical and psychological health of children~\cite{flood2009harms, quadara2017effects}.
Moreover, these toxic contents can even increase the number of criminal cases to a certain extent~\cite{Chen2020AutomaticDO}.
The numerous studies conducted on the topic have shown that the presence of toxic content poses a significant risk to social cohesion.

As a result, content moderation software that identifies, filters, blocks such content has become an area of immense interest for both academic researchers and industry professionals.
Toxic content detection has been widely formulated as a text classification task, and it has been tackled by various deep learning models, such as convolutional neuron networks, long-short-term-memory, and Transformers~\cite{Mishra2019TacklingOA, Schmidt2017ASO, Wu2018TwitterSD}.
Equipped with the advanced pre-trained language models (e.g., BERT~\cite{Devlin2019BERTPO} and RoBERTa~\cite{Liu2019RoBERTaAR}), industrial companies have recently gained substantiate improvement on the held-out accuracy of toxic content detection, providing commercial-level content moderation software such as Google~\cite{google2021}, Facebook~\cite{facebook2020}, Twitter~\cite{twitter2020}, and Baidu~\cite{notrobustbaidu}.

\begin{figure}[t]
    \centering
    \subfloat[Character-level Perturbation]{
        \includegraphics[width=\linewidth]{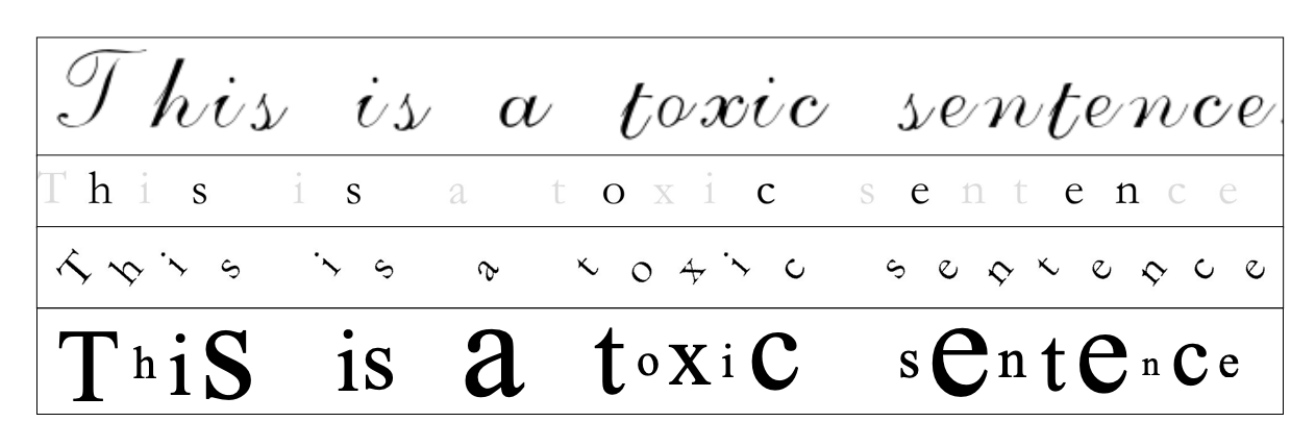}
    }
    \quad
    \subfloat[Paragraph-level Perturbation]{
          \includegraphics[width=\linewidth, height=.9\linewidth]{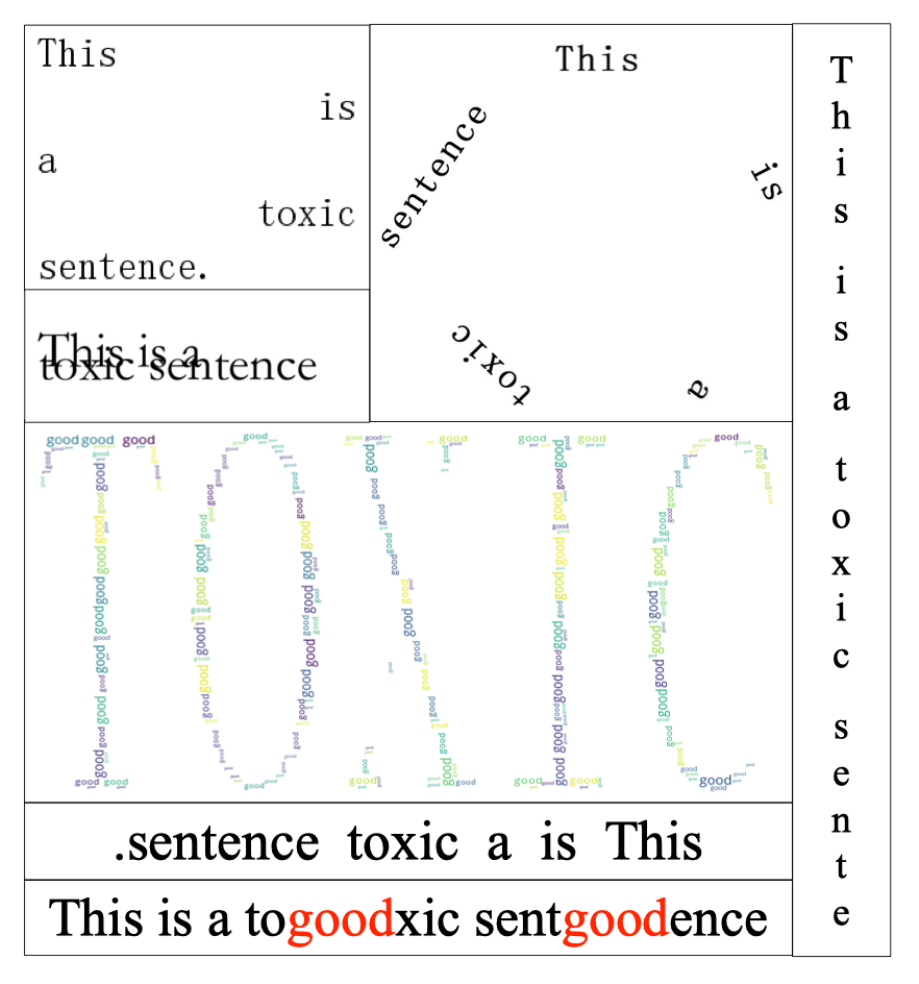}
          }
    \quad
    \subfloat[Picture-level Perturbation]{
          \includegraphics[width=\linewidth]{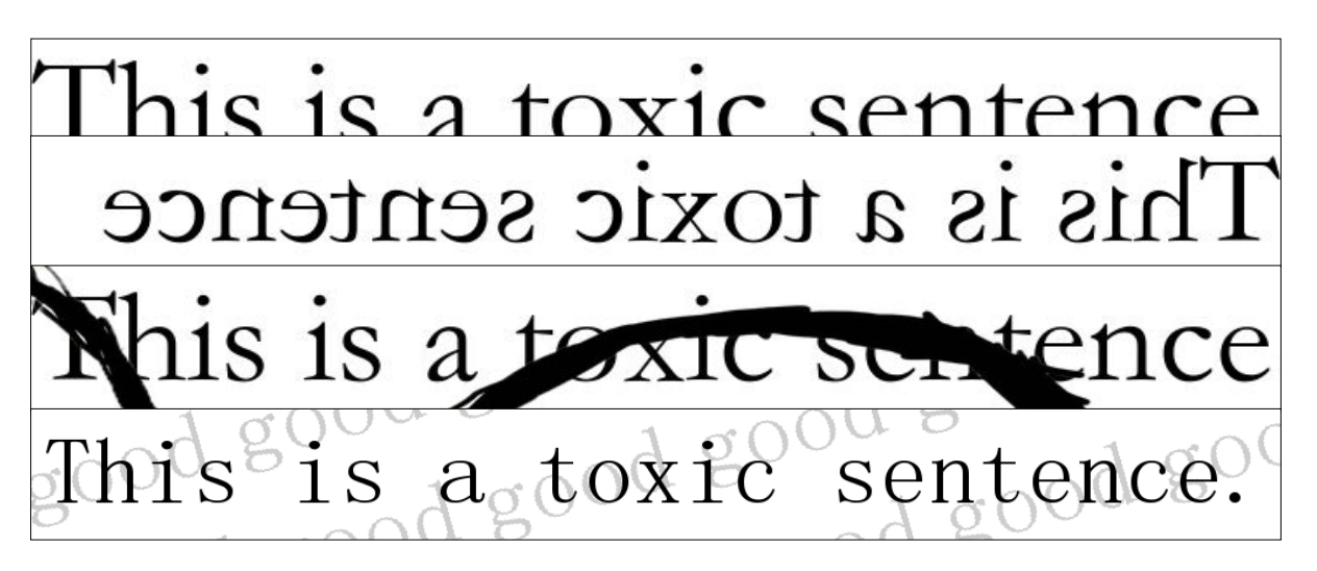}
          }
    \caption{Examples of pictures that contain toxic textual information with different perturbation methods : (a) Character-level perturbation, (b) Paragraph-level perturbation, and (c) Picture-level perturbation. Here we use a non-toxic seed "This is a toxic sentence" for demonstration.}
    \label{fig:example}
\end{figure}

Previous researchers have proposed several testing frameworks to measure the reliability of content moderation software, such as robustness against typos~\cite{Gao2018BlackBoxGO}, code-switch~\cite{Kapoor2019MindYL}, adversarial text attacks\cite{Li2019TextBuggerGA} and other human-intended perturbations \cite{Wang2023MTTMMT}.
However, existing testing work only focuses on textual perturbation, ignoring the spread of images.
In 2013, the number of images shared per day on Instagram and Facebook were 40 million and 300 million, respectively.
However, these numbers have significantly increased over the years, reaching 1.3 billion and 2.1 billion for Instagram and Facebook, respectively, by 2023~\cite{image-social-media}.
Malicious users can use images containing toxic texts to evade content moderation instead of simply using textual perturbations. The advantage of using images is that it provides  more degrees of freedom for adding perturbations, such as changing font, changing font color, changing font size, changing the way the text is arranged, rotation, and distortion, which cannot be easily adopted by previous perturbation methods for text~\cite{Li2019TextBuggerGA, Wang2023MTTMMT, Garg2020BAEBA, Jin2020IsBR}.
Figure~\ref{fig:example} shows some examples where the toxic information is hidden by different-level of perturbations, which are not covered by existing studies.

In this paper, we propose \methodname, a metamorphic testing framework to validate content moderation software, focusing on images containing toxic contents.
Specifically, to develop a comprehensive testing framework, we first need to understand how real-world malicious users evade moderation.
To this end, we conduct a pilot study (Section \ref{sec-mrs}) on 5,000 image messages collected from real users.
We study data in English and Chinese since English is the mostly used language for web contents~\cite{users-language} and China has the largest digital populations around the world~\cite{users-contry}.
We summarize 21 transformation rules, based on which we design metamorphic relations (MRs) across three perturbation levels: character level, paragraph level, and picture level.
Therefore, test cases generated under our MRs can reflect real-world user behaviors.

We then employ \methodname to validate moderation software with these generated test cases whose toxicity remains and can be easily recognized by humans.
In the evaluation, we apply \methodname to test five commercial content moderation software and a State-Of-The-Art (SOTA) moderation algorithm against three typical kinds of toxic content (\ie abusive language, malicious advertisement, and pornography). 
The results show that \methodname achieves up to $100\%$ Error Finding Rate (EFR) when testing commercial content moderation software provided by Google Cloud, Microsoft Azure, Baidu Cloud, Alibaba Cloud and Tencent Cloud, and the SOTA algorithm from the academy.
Additionally, we leverage the test cases generated by \methodname to retrain the model we explore, which largely improves model robustness (EFR of \methodname drops from $100\%$ to $6\%$) while maintaining the accuracy on the original test set. 
The main contributions of this paper are as follows:
\begin{itemize}[leftmargin=*]
    \item A preliminary study is conducted on 5,000 image messages in real-world scenarios, which results in a summary of 21 transform rules.
    \item Based on the rules, we introduce \methodname, the first comprehensive testing framework for textual toxic contents spread via images, which includes 21 metamorphic relations implemented in two languages: English and Chinese.
    \item Using \methodname, a thorough assessment is conducted on five content moderation software used in commerce and a SOTA academic model, revealing that toxic contents produced by \methodname could effortlessly evade moderation. Furthermore, we explore the feasibility of the generated toxic contents strengthening the robustness of the SOTA model.
\end{itemize}

The rest of the paper is organized as follows: We first introduce the background of metamorphic testing and content moderation in Section \ref{sec-backgound}; Then, in Section \ref{sec-mrs}, we introduce the design and implementation details of \methodname. In Section \ref{sec-experiment}, we conduct experiments to evaluate the effectiveness of \methodname; And in Section \ref{sec-discuss}, we summarize the contribution and analysis the threats to validity; Finally, we discuss the previous works that are related to ours in Section \ref{sec_related}.

\noindent \textbf{Content Warning}: We apologize that this paper includes instances of hostile, offensive, or explicit language for clarity, which have been quoted verbatim. Furthermore, in order to ensure the safety of our participants during the research, we take the following precautionary measures: (1) we consistently display a warning message to both the researchers and annotators at every stage, informing them that they could withdraw from the study at any time and (2) we offered psychological counseling after the study to alleviate any mental distress.

\section{Background}
\label{sec-backgound}

\subsection{Metamorphic Testing}

Metamorphic testing~\cite{Chen2020MetamorphicTA} has gained significant adoption as a testing technique for tackling the oracle problem.
Its fundamental concept involves identifying deviations in MRs across various software runs.
MRs describe the relationship between software input-output pairs under certain transformation rules.
Through metamorphic testing, a test case is modified by applying a pre-defined transformation rule, and the software's outputs for both the original and transformed test cases are compared to verify if they demonstrate the anticipated relationship.

In recent years, metamorphic testing has been applied to validate Artificial Intelligence (AI) software.
The objective of these endeavors is to automatically identify and report errors in AI software results using newly-defined MRs.
For example, Chen et al. \cite{Chen2008AnIA} investigated the use of metamorphic testing in bioinformatics applications.
Xie et al. \cite{Xie2011TestingAV} defined eleven MRs to test k-nearest neighbors and naive Bayes algorithms. 
Dwarakanath et al. \cite{Dwarakanath2018IdentifyingIB} presented eight MRs to test classifiers based on support vector machines and ResNet. 
Zhang et al. \cite{Zhang2018DeepRoadGM} tested autonomous driving systems by applying GANs to produce driving scenes with various weather conditions and checking the consistency of the system outputs.

\subsection{Content Moderation Software}

Commercial content moderation software has been implemented by several prominent companies such as Google~\cite{google2021}, Facebook~\cite{facebook2020}, Twitter~\cite{twitter2020}, and Baidu~\cite{notrobustbaidu} on their products.
Based on their official technical documents, the software typically employs a hybrid classification algorithm that combines neural network models and pre-defined rules.
This approach takes advantage of the strengths of both methods.
Neural network-based methods are proficient in understanding contextual and semantic information, whereas rule-based methods enable simple implementation of user-defined functionality.
Baidu, for instance, utilizes a deep neural network and an extensive list of pre-defined banned words to power its commercial content moderation software.

\section{\methodname}
\label{sec-mrs}

This section first introduces a pilot study on messages collected from real users (Section \ref{sec:pilotstudy}).
Then we introduce 21 metamorphic relations that are inspired by the pilot study.
These metamorphic relations can be grouped into three categories according to the perturbation performed: character-level perturbations (Sec.~\ref{sec:charlevel}), paragraph-level perturbations (Sec.~\ref{sec:paralevel}), and picture-level perturbations (Sec.~\ref{sec:picturelevel}).

\subsection{Pilot Study}
\label{sec:pilotstudy}

In this work, we intend to develop metamorphic relations that assume the seed test case (\ie a piece of text) and the generated test case (the picture) should have identical classification labels (\ie labeled as ``toxic content'') returned by the content moderation software. 
To generate effective test cases, we think the perturbations in our MRs should be:
\begin{itemize}[leftmargin=*]
    \item \textit{Semantic-preserving}: the perturbed test cases should have the identical semantic meaning as the seed.
    \item \textit{Realistic}: should reflect possible inputs from real users.
    \item \textit{Unambiguous}: should be defined clearly.
\end{itemize} 

In order to design satisfactory perturbations, we first conducted a pilot study on messages from real users to explore what kind of perturbations the users would apply to the toxic content to bypass the content moderation software.
We consider text messages from four platforms with a large number of users:
\begin{itemize}[leftmargin=*]
    \item Twitter. Twitter is an online social media and social networking service on which users post or reply to texts, images and videos known as "tweets". It has over 400 million monthly active users at the end of 2022.
    \item Instagram. Instagram is a photo and video sharing social networking service owned by Meta Platforms. It has over 2 billion monthly active users at the end of 2022.
    \item Sina Weibo. Sina Weibo is a Chinese microblogging website. It is one of the biggest social media platforms in China, with over 580 million monthly active users at the end of 2022.
    \item Baidu Tieba. Tieba is a Chinese online forum hosted by the Chinese web services company Baidu. It accumulated 45 million monthly active users and the number of its total registered users reached 1.5 billion at the end of 2021.

\end{itemize}

We collect $5,000$ images from the above website for manual inspection and recruited three annotators to label all the images independently.
All the annotators have a Bachelor's degree or above and are proficient in both English and Chinese.
Annotators were given extensive guidelines, test tasks, and training sessions on content moderation software and toxic content.
For each image, annotators were asked two questions. (1) Whether the image  is toxic or not? (2) Is the image intentionally perturbed to bypass the content moderation software?
After the annotation, we use the label that most workers agree with as the final human label and finally obtain 240 images that are labeled as ``toxic and intentionally perturbed'' images, which are used to design our perturbation methods.

\begin{table}
\caption{Summary of the perturbation categories in the pilot study.}
\label{tab:mrs}
\centering
\begin{tabular}{l l l}
\toprule
\bf Perturb Level & \bf Perturb Method  & \bf Percentage\\
\midrule
\multirow{5}{*}{Character Level}
& Font Change& 1.7\% \\
& Font Color Change & 2.9\%\\
& Font Size & 1.3\% \\
& Strikethrough & 3.8\% \\
& Char Rotation & 0.4\% \\
\midrule
\multirow{7}{*}{Paragraph Level}
& Circle & 0.4\%\\
& Vertical Direction & 0.4\%\\
& Right to left   & 14.9\%\\
& Align-left-then-right & 0.4\%\\
& Word Cloud & 1.3\%\\
& Overlap & 0.4\%\\
& Benign Text Camouflage  & 1.7\% \\
\midrule
\multirow{9}{*}{Picture Level}
& Blurring  & 14.6\% \\
& Crop  & 0.8\% \\
& Mirror  & 7.5\% \\
& Rotation  & 7.9\% \\
& Scribbling  & 46.3\% \\
& Distort  & 0.4\% \\
& Watermark  & 1.7\% \\
& To Gif  & 0.4\% \\
& Benign Image Camouflage  & 5.4\% \\
\bottomrule
\end{tabular}
\end{table}

We manually inspected all these toxic contents perturbed by the real users and collectively summarized 21 perturbation methods that real users have been using to evade moderation.
We categorize these toxic sentences from two perspectives: 1) the basic units of perturbation, such as character level (the perturbation methods that the malicious users can use when they are typing the characters), paragraph level (the perturbation methods that can use when typesetting the words into sentences or paragraph), and picture level (the perturbation methods that can be adopted after the malicious users screenshot the text to image); and 2) basic perturbation operation, such as change, insertion, deletion, split, and combination.
Accordingly, we derive 21 MRs based on 21 perturbation methods, where each MR assumes that the classification label returned by the content moderation software on the generated test case (i.e., images) should be the same as that on the seed (i.e., original text).
Table~\ref{tab:mrs} presents the 21 perturbation methods, their categories and the percentage of each in our study.
We will introduce the MRs (their corresponding perturbation methods) in the following.

\subsection{MRs with Character-Level Perturbations}
\label{sec:charlevel}

The character-level perturbations are the perturbation methods that malicious users can use when they are typing the characters.

\noindent \textbf{MR1-1 Font Change}

This MR is selecting a font that makes the text hard to be recognized for content moderation software. And when the toxic words are not recognized, the image will have more possibilities to bypass the moderation.
For example, Cursive is a style of penmanship in which characters are written joined in a flowing manner, in contrast to block letters. Changing the font from Times New Roman to Cursive will increase the difficulty of character recognition.

To implement this MR, we adopt the semi-cursive script font\footnote{https://www.fonts.net.cn/font-34032272562.html} and use the Python Imaging Library(PIL)\footnote{https://pypi.org/project/Pillow/} to change the font in the image with  ImageFont.truetype() function.

\noindent \textbf{MR1-2 Font Color Change}

This MR is using a font color, such as a color that is close to the background color, so as to make it harder for content moderation software to recognize the text. 

To implement this MR, we adopt a Python library named Pygame\footnote{https://www.pygame.org/} to render the text and use Pygame.font.render() function to control the color of each character. 

\noindent \textbf{MR1-3 Font Size Changing}

This MR's perturbation method is using different font sizes for different characters or words, aiming to make it more difficult for software to recognize the whole text. For example, most of the characters are in 24px but the toxic words or some of their characters are intentionally set to 4px, aiming to make the moderation software overlook the small toxic words. If so, there may have more possibilities to bypass the moderation.

To implement this MR, we adopt the python library Pygame. We use Pygame.font.render() function to control the size of different characters.

\noindent \textbf{MR1-4 Strikethrough}

This MR perturbs texts by adding horizontal lines through their center, aiming to make it more difficult for software to recognize the text. 

To implement this MR, we can use PIL to open the image of text and add two horizontal lines to it. The first line is located at 33\% of the image height, and the second is located at 66\%.

\noindent \textbf{MR1-5 Character Rotation}

This MR rotates each character by a random angle, aiming to make it more difficult for software to recognize the text. 

To implement this MR, the Pygame library was used. Each character is rotated individually by a random degree  with pygame.transform.rotate() function.

\subsection{MRs with Paragraph-Level Perturbations}
\label{sec:paralevel}

The paragraph-level perturbations are the perturbation methods that malicious users can use when typesetting characters or words into sentences or paragraphs. Different from character-Level perturbations, paragraph-level perturbations do not make modifications to any characters.

\noindent \textbf{MR2-1 Circle}

Traditionally, the characters or words  are typeset in a line. This MR layout the text in a circle, aiming to make it more difficult for software to understand the meaning of the text. 

To implement this MR, both the Python math and Pygame libraries were used. We first use the math library to calculate the position of each character. Then we place each character in its appropriate position using the blit() function.

\noindent \textbf{MR2-2 Vertical Direction}

Traditionally, the characters or words are typeset in a left-to-right manner. This MR typeset the characters or words vertically in a top-to-bottom manner, aiming to make it more difficult for software to understand the meaning of the text. 

To implement this MR, we combine each image vertically with a python library NumPy\footnote{https://pypi.org/project/numpy/}. We first use PIL to write each character or word in a small image. 
 Then we concatenate these small images in an up-to-bottom manner with NumPy array manipulation.

\noindent \textbf{MR2-3 Right-to-left}

Traditionally, the characters or words are typeset in a left-to-right manner. 
 This MR typeset the characters in a word or the words in a sentence in a right-to-left manner, aiming to make it more difficult for software to understand the meaning of the text. 

To implement this MR, we first reverse the order of the sentences. Then we place the reversed sentence in the image by blit() function in Pygame.

\noindent \textbf{MR2-4 Align-left-then-right}

This MR splits a sentence into multiple lines and typesets the first line in a manner of align-left, then the second line align-right, the third line align-left, ..., aiming to make it more difficult for software to understand the meaning of the text. 

To implement this MR, we also use PIL to write each character or word in a line. Different from MR2-2 which shows the character or word in the middle, this time we arrange the first character or word on the left, and arrange the second one on the right, ending up with a left-right-left-right manner. Then we use the NumPy array  to concatenate them vertically.

\noindent \textbf{MR2-5 Word Cloud}

This MR is using many small benign words to fill the outline of large characters or words, aiming to make it more difficult for software to recognize the large characters or words. For example, many small "good"  forming a big "bad" shape could be recognized as many "good" rather than the big "bad" for content moderation software. 

To implement this MR, we use the Python library WordCloud\footnote{https://pypi.org/project/wordcloud/}, which is able to generate word clouds according to the shape we specify. We first use the PIL library to write the text as the target shape. Then we pass it into the WordCloud() function to get the word cloud image.

\noindent \textbf{MR2-6 Overlap}

This MR reduces the line spacing or word spacing to make some overlap between words or characters, so as to make it more difficult for software to recognize the words. 

To implement this MR, we arrange each word in a sentence to the image in a partial overlap manner with blit() function in the Pygame library.

\noindent \textbf{MR2-7 Benign Text Camouflage}

This MR is hiding the words or characters in plenty of benign words as context and uses some way to highlight the original  characters or words, such as using a different font color or circling them in red. Human can understand the toxic nature of such content by only paying attention to the highlighted characters or words but it will be much harder for software to understand that.

To implement this MR, we first generate plenty of non-toxic words or characters and insert them between or surrounding the words of the original toxic sentence as “ benign text padding”. Then, we make a list to record the position of the original toxic words we want to hide and use PIL to draw red circles on the image to highlight each hidden word.

\subsection{MRs with Picture-Level Perturbations}
\label{sec:picturelevel}

The picture-level perturbations are the perturbation methods that can be adopted after the malicious users screenshot the text to the image. Different from the perturbation methods introduced above, picture-level perturbations only perturb the images.

\noindent \textbf{MR3-1 Blurring}

This MR blurs the picture or reduces the resolution of the image, aiming to make it more difficult for software to recognize the words in the picture. 

To implement this MR, we adopt the Python libraries cv2. We first generate an image with a toxic sentence with Pygame. Then we blur that image with the cv2.blur() function.

\noindent \textbf{MR3-2 Crop}

This MR crops the image so that only part of each character in the image is preserved, aiming to make it more difficult for software to recognize the original words. But the human can easily recognize the character by imagining what the whole character or words look like.

To implement this MR, we adopt the Python library PIL. We first use PIL to generate an image with the toxic sentence. Then we crop the bottom 30\% of each image with the crop() function. 

\noindent \textbf{MR3-3 Mirror}

This MR mirrors the picture, aiming to make it more difficult for software to recognize the words. The mirrored texts are also hard for humans to understand, so the users need to mirror the picture back to read the message, such as using photo editing software on their cellphones or personal computer.

To implement this MR, we first generate an image with the toxic sentence and then mirror the image with the transpose() function.

\noindent \textbf{MR3-4 Rotation}

This MR rotates the picture at an angle, such as $90^{\circ}$ or $180^{\circ}$,  aiming to make it more difficult for software to recognize the words. Human can easily understand the content by either rotating back the image using photo editing software or rotating their cellphones.

To implement this MR, we first create the image with the toxic sentence and then rotate it 45 degrees with pygame.transform.rotate() function. 

\noindent \textbf{MR3-5 Scribbling}

This MR adds meaningless marks or lines, with a pencil or pen, to the picture, aiming to make it more difficult for software to recognize the words. But as humans, we can easily or even automatically ignore the scribbling and understand the content in the original image.

To implement this MR, we first prepare a scribbling image, which contains human-intended scribbling. Then we generate an image for each toxic sentence by Pygame and resize the scribbling image with PIL's resize() function according to the size of the image with the toxic sentence. Finally, we superimpose the resized scribbling image and the image with a toxic sentence.

\noindent \textbf{MR3-6 Distort}

This MR makes a non-rectilinear projection of the image and can significantly disrupt the image's quality. For example, an image's straight lines will appear deformed or curved unnaturally. 

To implement this MR, we use the function resize() in the PIL library. We change the parameters in resize() to distort the  image, such as stretching and bending.

\noindent \textbf{MR3-6 Watermark}

This MR intentionally superimposes plenty of logo, text, or pattern onto the image. Its purpose is to make it more difficult for software to recognize the original messages in the image, since the software may pay more attention to the watermark, or the original toxic content is occluded by the watermark.

To implement this MR, we use the PIL library. We first make the watermark image by initializing a grayscale image
and then inserting plenty of "good words" to fill the image. After that, we rotate the watermark image and superimpose it with the original image.

\noindent \textbf{MR3-7 To Gif}

GIF (the Graphics Interchange Format) is an image type that contains many frames and allows a separate display for each frame. This MR converts the image to GIF by displaying a portion of the message in one frame and displaying the other in another frame. When the frames per second are more than 90 Hz, humans watching this GIF is like looking at a single picture. While for computer software, each moment can only capture a portion of the message, which may lead to the bypassing of moderation.

To implement this MR, we use the PIL library to generate two images: one shows the odd-bit characters and another shows the even-bit characters, and each uses empty space as padding. Finally, we use the save() function in PIL and set the format='GIF' to get the gif we want.

\noindent \textbf{MR3-8 Benign Images Camouflage}

This MR hides the toxic picture within many benign images. 
For example, a malicious advertisement can be surrounded by two unrelated and non-commercial images, generating a long picture, which may bypass the malicious advertisement detection model.

To implement this MR, we first randomly download some landscape photos from the Internet as benign images. Then use the NumPy library to combine the benign images with the toxic image by using numpy.atleast\_2d() and numpy.append() functions.

\subsection{Discussion}

\noindent \textbf{Combinations of Different MRs.}
According to our pilot study, we find that some of the user input involved multiple perturbations. And we can use a combination of different MRs to generate diverse test cases.
However, to control the experimental variables and the test cases' readability, we only utilize one MR in each test case.
We evaluate the impact of MR combinations in Section \ref{subsec-testing-software-with-tool}.

\noindent \textbf{Generalization to other software and languages.}
In this work, we focus on content moderation software and implement our MRs for the two most widely used languages: English and Chinese.
However, based on our design methodology, these MRs can be easily generalized to other languages and to test other NLP software, such as software for user review analysis and machine translation.

\section{Evaluation}
\label{sec-experiment}

\begin{table}
\centering
\caption{Statistics of Toxic Datasets.}
\begin{tabular}{l r r r r}
\toprule
\bf Dataset & \bf \#Sent & \bf Lang &  \bf Type & \bf Source\\
\midrule
HateOffensive & 24.8K  &  English & Abuse   &  Twitter \\
Dirty  &  2.5K & Chinese & Abuse & Weibo \\
SMSSpam & 5.5k & English &  Spam & Grumbletext \\
SpamMessage & 60K & Chinese & Spam & Taobao\\
Sexting  & 0.5K & English & Porno & Github \\
Midu  & 7.3K & Chinese & Porno & Midu \\
\bottomrule
\end{tabular}
\label{tab:data-statistics}
\end{table}


To evaluate the effectiveness of \methodname, we use our method to test three commercial software products and two SOTA algorithms for content moderation.
In this section, we try to answer the following three Research Questions (RQs):
\begin{itemize}[leftmargin=*]
    \item RQ1: Are the test cases generated by \methodname toxic and realistic?
    \item RQ2: Can \methodname find erroneous outputs returned by content moderation software?
    \item RQ3: Can we utilize the test cases generated by \methodname to improve the performance of content moderation?
\end{itemize}

\subsection{Experimental Settings}

\subsubsection{Datasets}

We used different kinds of datasets as seed data to validate \methodname.
Previous researchers have collected, labeled, and released various types of data for research purposes.
In this paper, we choose the datasets with the highest citations according to Google Scholar or those with the most stars on GitHub. Important statistics of the six datasets are shown in Table~\ref{tab:data-statistics}.

\begin{itemize}[leftmargin=*]
    \item HateOffensive\footnote{https://github.com/t-davidson/hate-speech-and-offensive-language} \cite{hateoffensive} is a GitHub repository containing 24,802 English hate speech sentences collected from Twitter. 
    \item Dirty\footnote{https://github.com/pokemonchw/Dirty} is a GitHub repository containing 2.5k Chinese toxic sentences with abusive and sexual words.
    \item SMS Spam Collection\footnote{https://www.kaggle.com/uciml/sms-spam-collection-dataset} is a set of tagged SMS messages. It contains 5,574 SMS messages in English, tagged as being ham (legitimate) or spam. The data was manually extracted from the Grumbletext website, a UK forum in which cell phone users make public claims about SMS spam messages.
    \item SpamMessage\footnote{https://github.com/hrwhisper/SpamMessage} is a Github repository containing 60k malicious advertisement messages in Chinese.
    \item Sexting\footnote{https://github.com/mathigatti/sexting-dataset} is an English pornographic text dataset containing 537 sexual texting messages.
    \item Midu \cite{Song2021EvidenceAN} is a Chinese novel paragraph dataset collected from an online literature reading platform called MiDu App\footnote{http://www.midureader.com/}. It is a corpus with 62,876 paragraphs including 7,360 pornographic paragraphs and 55,516 normal paragraphs.
\end{itemize}

\subsubsection{Software and Models Under Test}
\label{subsec:models}

In the implementation, we test 5 commercial software products provided by large Internet companies, \ie Google Cloud Vision \footnote{https://cloud.google.com/vision/docs/detecting-safe-search}, Microsoft Azure Image Moderation\footnote{https://learn.microsoft.com/en-us/azure/cognitive-services/content-moderator/image-moderation-api}, Baidu Cloud Content Moderation \footnote{https://cloud.baidu.com/doc/ANTIPORN/s/6ki012lqu}, Tencent Cloud Image Auditing\footnote{https://cloud.tencent.com/document/product/1235} and Alibaba Cloud Content Safety\footnote{https://help.aliyun.com/document\_detail/70409.html}, all of which are the official content moderation software from big technology companies with more than 100 millions of users. In particular, all the software products are the latest version by March. 1st, 2023, when the experiments were conducted. The version information of the software under test is listed in Table~\ref{tab:software_version_info}.
Besides commercial software products, we also test popular research models. Since there is no end-to-end open-sourced model or publicly available benchmark for toxic screenshot detection, we follow \cite{Vidgen2019ChallengesAF} to use an optical
character recognition (OCR) + text classification pipeline to detect multi-modal toxicity. Specifically, we first adopt Tesseract OCR \footnote{https://github.com/tesseract-ocr/tesseract}, the most famous open-source OCR toolkit in Github with 50K stars, to extract the textual information from the image. Then we adopt open-sourced text classification models to detect the toxicity from the extracted text. we select models from GitHub and Huggingface Model Zoo\footnote{https://huggingface.co/models} with the highest downloads and stars in recent three years.
For abuse detection, we select HateXplain \cite{Mathew2021HateXplainAB}, a BERT model fine-tuned on abuse detection datasets.
Since there are no publicly available spam and pornography detection models, we do not test these research models in our experiments.

\begin{table}
\centering
\caption{Software Version Information.}
\begin{tabular}{l | r  r }
\toprule
\bf Software & \bf Version & \bf  Lanch Date   \\
\midrule
Google& builtin/latest & 2022.12.16 \\
Microsoft& 2023.02.15  & 2023.02.15 \\
Baidu &  4.16.3 & 2022.03.25 \\
Tencent&  2022-07-27 & 2022.07.27\\
Alibaba & 2022.06.15  & 2022.06.15  \\
\bottomrule
\end{tabular}
\label{tab:software_version_info}
\vspace{-12pt}
\end{table}

\subsection{RQ1: Are the test cases generated by \methodname toxic and realistic?}

\methodname aims to generate test cases that are toxic and are as realistic as the ones real-world users produce to evade moderation.
Thus, in this section, we evaluate whether the generated test cases are still toxic (i.e., semantic-preserving) and whether they are realistic.
We conduct human annotation via crowd-sourcing. We first generated $10$ images with each perturbation method, ending up with 210 test cases for annotation.  

For each images, we asked the following two questions:
(1) From ``$1$ strongly disagree'' to ``$5$ strongly agree'', how much do you regard the image as toxic content (abuse, pornographic, or spam)? 
(2) From ``$1$ strongly disagree'' to ``$5$ strongly agree'', how much do you think the perturbation is realistic in the sense that real users may use it? 

We distribute the questionnaire and recruit 20 crowd workers on Tencent Wenjuan\footnote{https://wj.qq.com/} with Bachelor’s degrees
or above and proficiency in both English and Chinese. Before annotation, we provide instructions about the type of questions and asked them to make subjective judgments in the annotation. We do not provide additional training to avoid potential bias from us. 

Annotation results show that: 1) the generated test cases are toxic, with an average score of 4.46; and 2) the generated test cases are realistic, with an average  score of 4.05. We followed \cite{Kirk2021HatemojiAT} to measure the inter-annotator agreement using Randolph’s Kappa, obtaining a value of 0.81 for the test cases, which indicates "almost perfect agreement". 

\begin{tcolorbox}[width=\linewidth, boxrule=0pt, colback=gray!20, colframe=gray!20]
\textbf{Answer to RQ1:}
The test cases generated by \methodname are toxic and realistic.
\end{tcolorbox}

\subsection{RQ2: Can \methodname find erroneous outputs returned by content moderation software?}
\label{subsec-testing-software-with-tool}

\begin{table*}
\centering
\caption{Error Finding Rates of commercial content moderation software, including Google Cloud(G), Microsoft Azure (M), Alibaba Cloud (A), Baidu Cloud (B) and Tencent Cloud (T), and Academic Models (AM).}
\begin{tabular}{l l  | l l l l l l| l l l |l l l l l }
\toprule
\multirow{2}{*}{\bf Level} & \multirow{2}{*}{\bf Perturbation Methods } & \multicolumn{6}{c}{\bf Abuse Detection}   & \multicolumn{3}{c}{\bf Spam Detection}  & \multicolumn{5}{c}{\bf Pron Detection}\\
\cmidrule(lr){3-8} \cmidrule(lr){9-11}  \cmidrule(lr){12-16}  
& &  \bf G &  \bf M &\bf A  &\bf B & \bf T & \bf AM  &  \bf A  &\bf B  & \bf T   & \bf G & \bf  M & \bf A  &\bf B & \bf T \\
\midrule
\multirow{5}{*}{Char}&  Font Change & 2 & 4 & 14 & 30 & 6 &  100 &34 & 26 & 44 &  0 & 2 & 34 & 18 & 12   \\
& Font Color  & 4 & 40 & 4 & 8 & 0 & 86 & 8 & 0 & 12 &   4 & 40 & 20 & 4 & 0  \\
& Font Size &  36 & 70 & 20 & 36 & 8 & 88 &  66 & 62 & 58 &  46 & 84 & 30 & 20 & 0  \\
& Strikethrough   &  78 & 50 & 92 & 0 & 100 & 100 & 96 & 0 & 100 &    84 & 30 & 88 & 4 & 100   \\
& Char Rotation  & 100 & 100 &100 &100 &48 & 100 &100 &98 &86  &  100 & 96 &100 &100 &24  \\
\midrule
\multirow{7}{*}{Paragraph}& Circle  &  8 & 4 & 96 & 92 & 20 & 82 & 
  100  &  98 & 72 &    4 & 4 & 98 & 94 & 8  \\
& Vertical Direction & 72 & 100 &32 &6 &2 & 44 & 100  &  66 &2 &  68 & 100 &60 &36 &0\\
& Right to left   & 8 & 4 & 48 & 22 & 26 & 58  & 58 & 98 & 88 &  14 & 2  & 64 & 40 & 42   \\
& Align-left-then-right  & 98 & 100 & 36 & 100 & 0 & 48  &
 86 & 100 &  0 &    96 & 100 & 28 & 100 & 0  \\
& Word Cloud  & 96 & 18 & 100 & 100 & 100 & 100 &  98 & 100 & 100 &    92 & 2 & 100 & 100 & 100   \\
& Overlap  & 84 & 64 &90 &88 &74 & 100 &100 &100 &100  &  92 & 70 &96 &90 &94   \\
& Benign Text Camouflage  & 98 & 94 &98 &2 &48 &  100 & 100 &96 &100 & 100 & 94 &100 &6 &34  \\
\midrule
\multirow{9}{*}{Picture} &  Blurring & 90 & 100 & 94 & 100 & 100 & 100  & 86 & 84 & 100 &    94 & 100 & 98 &  80 & 100 \\
&  Crop & 22 & 44 & 96 & 96 &  74 & 100 & 100 &  96 & 100 &   12 & 30 & 98 & 94 & 94   \\
&  Mirror & 100 & 86 & 98 & 100 & 100 & 100 & 100 & 100 & 100 &   100 & 88 & 100 & 100 & 100  \\
&  Rotation &  0 & 2 & 16 & 100 & 18 & 100 & 22 & 100 & 100 & 0 & 0 & 2 & 100 & 18  \\
&  Scribbling & 6 & 18 & 34 & 34 & 12 & 76 & 36 & 10 & 68 &  4 & 18 & 44 & 26 & 14  \\
&  Distort & 2 & 12 & 94 & 100 & 42 & 88 & 100 & 100 & 66 &   2 & 4 & 100 & 100 & 58  \\
&  Watermark & 70 & 4 & 12 & 6 & 32 & 60 & 32 & 10 & 88 &    72 & 0 & 40 & 14 & 32   \\
&   To Gif  & 96 & 96 & 56 & 66 & 40 & 100 & 96 & 100 & 100 &   100 &  96 & 46 & 82 & 52  \\
&  Benign Image Camouflage  &  100 & 0 & 72 & 10 & 6 &  46  & 100 & 6 & 0 & 100 & 0 & 54 & 14 & 0    \\
\midrule
Multi & Perturb Combinations & 58 & 68 &94  &98  &34  & 100  & 100  &100  &70  &   78 & 86  &94  &100  &36   \\
\bottomrule
\end{tabular}
\label{tab:abuse}
\end{table*}

\methodname aims to automatically generate test cases to find potential bugs in current content moderation software.
Hence, in this section, we evaluate the number of bugs that \methodname can find in the outputs of commercial content moderation software and academic models.
We first input all the original sentences and obtain the classification label for each software product or model under test.
If an original sentence was labeled as ``non-toxic'', it would be filtered out because we intend to find toxic contents that can evade moderation.
The remaining sentences will be regarded as seed sentences for test case generation.
Then, we conduct perturbations in \methodname's MRs on the seed sentences to generate test cases. 
Finally, we use the generated test cases to validate the software products and academic models. 
In particular, we check whether these test cases were labeled as ``toxic'' or ``non-toxic''.
Since the generated text should preserve the semantics of the seed sentence, they are supposed to be labeled as ``toxic''. If not, the generated test cases evade the moderation of the software products or academic models, indicating erroneous outputs.
To evaluate how well \methodname does on generating test cases that trigger errors, we calculate Error Finding Rate (EFR), which is defined as follows:

$$\text{EFR} = \frac{\text{the number of misclassified test cases}}{\text{the number of generated test cases}} * 100\%.$$

The EFR results are shown in Table~\ref{tab:abuse}. 
In general, \methodname achieves high EFRs.
Using different MRs, \methodname achieves up to $100\%$ EFR when testing moderation software provided by Google, Microsoft, Baidu, and Tencent and Alibaba, and it obtains up to $100\%$ EFR when testing the SOTA academic models.

One common concern about AI software testing is whether the software performs well on existing test cases, which are toxic inputs for content moderation. To address this concern, we conduct a lightweight experiment to evaluate the effectiveness of the software under test in detecting toxic contents from the Internet. Since there is no publicly-available toxic (hateful, porno, and malicious ad) image benchmark dataset (probably due to the toxic nature), we manually collect a dataset with 50 hateful images, 50 porno images, and 50 ad images from the Internet. The average detection rate of five content moderation software is 97.8\%, indicating the effectiveness of the software. Thus, we think the high EFR achieved by \methodname is exciting.

\begin{tcolorbox}[width=\linewidth, boxrule=0pt, colback=gray!20, colframe=gray!20]
\textbf{Answer to RQ2:}
\methodname achieves up to $100\%$ EFR when testing moderation software provided by Google, Microsoft, Baidu, and Tencent and Alibaba, and it obtains up to $100\%$ EFR when testing the SOTA academic models.
\end{tcolorbox}

\subsection{RQ3: Can we utilize the test cases generated by \methodname to improve the performance of content moderation?}

We have demonstrated that \methodname can generate toxic and realistic test cases that can evade the moderation of commercial software products and SOTA academic models.
As shown in the ``Abuse Detection'' column in Table~\ref{tab:abuse}, \methodname achieves high EFR on academic models for most of its MRs (e.g., $100\%$ for MR1-1 Font Change), indicating the generated test cases can easily fool the models.
The following substantial question is: can these test cases be utilized to improve the performance of content moderation?
In other words, we hope to improve model robustness.
A natural thought is to retrain the models using test cases generated by \methodname and check whether the retrained models are more robust to various perturbations.

Specifically, we select the Abuse Detection task and use the Hate-Offensive Dataset~\cite{hateoffensive}.
We split the dataset into three parts: training set, validation set, and test set with the ratio of $6$:$2$:$2$.
We first fine-tune a pre-trained BERT model~\cite{Devlin2019BERTPO} on the training set as our abuse detection model, which is a widely used scheme for text classification.
We adopt the default fine-tuning settings suggested by Huggingface\footnote{\url{https://huggingface.co/transformers/v3.2.0/custom_datasets.html}}.
Specifically, we train the model with $3$ epochs, a learning rate of $5\times 10^{-5}$, a batch size of $16$, $500$ warming up steps, and a weight decay of $0.01$. 
We select the model with the highest accuracy on the validation set and use \methodname to test its robustness.

Then, for retraining with \methodname, we conduct fine-tuning with the failed test cases generated by \methodname.
We generated test cases with \methodname and randomly collected $1000$ cases that could fool the model. We first use the (image, text) pairs to retrain the LSTM-based ORC model in tesseract, following its official document \footnote{https://github.com/tesseract-ocr/tesstrain/wiki}. Then we label all the text as toxic content and add the (text, label) pairs to the original training set to retrain the BERT-based abuse detection model.
The setting of hyper-parameters is identical to that of regular training mentioned above.

\begin{table}
\centering
\caption{Error Finding Rates (EFRs) on abusive language detection models after retraining on the original test set and the test cases generated by \methodname.}
\centering
\normalsize
\begin{tabular}{l l l | l }
\toprule
\bf Level & \bf Perturb Methods & \bf Ori  &\bf Aug  \\
\midrule
\multirow{5}{*}{Char}& Font Change & 100 & 10 \\
& Font Color & 86 & 8 \\
& Font Size &  88  & 12 \\
& Strikethrough & 100 & 12 \\
& Char Rotation &   100 & 6 \\
\midrule
\multirow{7}{*}{Para}& Circle & 82 & 16 \\
& Vertical Direction & 44 & 4 \\
& Right to left & 58 & 4 \\
& Align-left-then-right & 48  & 8 \\
& Word Cloud & 100 & 16 \\
& Overlap & 100 & 18 \\
& Benign Text  & 100 & 28 \\
\midrule
\multirow{8}{*}{Picture} &  Blurring  & 100 & 24 \\
& Crop & 100 & 18 \\
& Mirror & 100 & 36 \\
& Rotation & 100 & 14 \\
& Scribbling & 76 & 16 \\
& Distort & 88 & 12 \\
& Watermark & 60 & 8 \\
& Benign Image  &  46 & 4 \\
\bottomrule
\end{tabular}
\label{tab:abuse_improve}
\end{table}

To validate the effectiveness of robust retraining with \methodname, we use \methodname to test the model after robust retraining, denoted as ``Aug'', and compared the EFRs with the original model's, denoted as ``Ori''.
The results are presented in Table~\ref{tab:abuse_improve}.
We can observe that the test case generated by \methodname can largely improve the robustness of the content moderation models in the sense that the EFRs have been significantly reduced (e.g., from $100\%$ to $10\%$ for the MR1-1 Font Change).
In other words, after retraining with \methodname's test cases, the model is rarely fooled by all the perturbations.
Moreover, the model's accuracy remains on par after robust training, which means the retraining did not affect model performance on the original test set.

We do not conduct experiments on improving industrial models because industrial moderation only provides APIs while robust retraining requires access to model internals. However, we believe robust retraining with \methodname’s test cases would also improve the robustness of industrial models because the underlying models are similar. In the future, we can study how to improve the robustness of industrial moderation by designing a preprocessing module to detect and filter out/reverse-perturb intentionally-perturbed inputs.

\begin{tcolorbox}[width=\linewidth, boxrule=0pt, colback=gray!20, colframe=gray!20]
\textbf{Answer to RQ3:}
Test cases generated by \methodname can effectively improve the robustness of academic content moderation models.
\end{tcolorbox}

\section{Discussion}
\label{sec-discuss}

\subsection{Compared with AI Testing and Adversarial Attack Methods}

In this section, we will illustrate the difference and advantages of \methodname compared to other AI testing and adversarial attack methods, which are also aiming to find the error in image-input software.

First, \methodname is in a purely black-box setting while most of the other methods are not. The majority of AI testing~\cite{Tian2018DeepTestAT,Pei2017DeepXploreAW, Guo2018DLFuzzDF,Xie2019DeepHunterAC,Guo2018DLFuzzDF, Ma2019DeepCTTC, Ma2018DeepGaugeMT} and adversarial attack methods~\cite{Biggio2013EvasionAA, Szegedy2013IntriguingPO, MoosaviDezfooli2015DeepFoolAS, Carlini2016TowardsET} are in a white-box setting where the deployed model, e.g., inputs, model architecture, and specific model parameter values, are known. These works utilize the neuron coverage or gradient to generate or select test cases. Another thread of work~\cite{Zhang2019DeepSearchAS, Chen2017ZOOZO} is in a black-box setting, where not only the output labels but also the confidence scores are needed. However, both of the above settings are not practical in testing content moderation software, where neither the model details nor the confidence scores are provided.

Second, \methodname is more comprehensive than previous black-box testing and adversarial attack methods. 
For example, DeepTest~\cite{Tian2018DeepTestAT} also adopted some picture-level perturbations, such as Crop, Rotation, Blurring, and Distort, to test the Neural-Network-based Autonomous Cars. DeepRoad~\cite{Zhang2018DeepRoadGM} applied the Generative Adversarial Networks to generate different real-world weather scenes. Kurakin et al.~\cite{Kurakin2016AdversarialEI} found that the adversarial examples in physical world scenarios, such as  printouts of photos or cropped photos, can fool the image classification models. Athalye et al.~\cite{Athalye2017SynthesizingRA} synthesized real-world adversarial examples by rescaling, rotation, lightening or  darkening the original photos.
Other query-based adversarial attack methods~\cite{MoosaviDezfooli2016UniversalAP, Zhou2018TransferableAP} aims to generate invisible perturbations to input images. While \methodname contains 21 kinds of perturbations summarized from our pilot study on real user inputs.

\subsection{Threats to Validity}

The validity of our study may be subject to some threats.
\textit{The first threat} is that the test cases generated by \methodname after many perturbations may become ``non-toxic'', leading to false positives.
To relieve this threat, we conducted a user study to validate whether the generated test cases are toxic or not.
We further asked the annotators to label whether the test cases reflect inputs from real users. The results show that the generated test cases are toxic and realistic.
\textit{The second threat} is whether the content moderation software products we test are good and representative of what the industry uses. To relieve this issue, we evaluated the effectiveness of these products. The average detection rate of five content moderation software products is 97.8\%, indicating their effectiveness. As for representativeness, all five commercial software products are paid cloud services provided by the could platform from the big companies. Moderation services and other cloud services have become an important source of income for them. Meanwhile, a huge amount of downstream companies and users are using paid services provided by these companies. The customer list can be seen from Google\footnote{https://cloud.google.com/customers\#/products=Data\_Analytics}, Amazon\footnote{https://aws.amazon.com/machine-learning/customers/} and Baidu\footnote{
https://cloud.baidu.com/partner/plan.html\#search}. Thus, we believe the moderation software studied in our paper has a significant impact on real users and is representative of industry practice.
\textit{The third threat} is that our \methodname could be outdated with the bypass techniques evolving.
To reduce this threat, we provide a comprehensive workflow: study the user behaviors, summarize and design the MRs, generate test cases, and use failure cases to improve the robustness.
If other bypass techniques were proposed, people could follow this workflow to design new MRs.
We also believe that automated MR generation is a promising and useful direction. This line of research mainly focuses on automated generation of a specific kind of MRs (e.g., polynomial MRs~\cite{Zhang2014SearchbasedIO, Zhang2019AutomaticDA} or automated MR generation leveraging software redundancy~\cite{Carzaniga2014CrosscheckingOF}. Since automated MR generation for content moderation software faces different challenges, we regard it as an important future work.

\section{Related Work}
\label{sec_related}

\subsection{Testing AI Software}

\textit{AI software} has enabled a diversity of applications, for example, autonomous driving and face recognition.
Nevertheless, AI-based models are known to be of low robustness, which can generate undesired outputs causing severe accidents \cite{notrobustself-driving, notrobusttesla}.
Researchers have designed diverse approaches to generating adversarial examples or test cases that can fool AI software~\cite{Carlini2016HiddenVC, Pham2021DEVIATEAD, Wang2021RobOTRT, Zhang2022ImprovingAT, Zhang2023ImprovingTT}.
At the same time, researchers have also proposed algorithms to improve AI software’s robustness, for example, the robust training mechanism~\cite{Madry2018TowardsDL, Asyrofi2021CanDT, Gao2020FuzzTB, Wang2023ValidatingMC} and network debugging~\cite{Ma2018MODEAN, Tao2020TRADERTD}.

\textit{NLP software} has become a major application of AI software in recent years.
Sentiment analysis~\cite{Zhang2017SentimentAA, Wang2017EmotionRW} enables online review recommendation;
Sequence-to-sequence models make substantial progress in machine translation~\cite{Bahdanau2015NeuralMT, Wang2022UnderstandingAI, Jiao2022TencentsMM} and text-to-speech synthesis~\cite{Wang2017TacotronTE, Ma2018FPETSFP}. 
Both software engineering and NLP researchers have started to explore the robustness of NLP software~\cite{Gupta2020MachineTT, He2021TestingMT, Jiao2023IsCA}.
Specifically, deep learning models can help generate test cases for NLP software~\cite{Li2020BERTATTACKAA}.
Machine translation can be tested under a word-replacement-based approach~\cite{Sun2022ImprovingMT}.
Other NLP software, such as sentiment analysis, duplicate question answering, and machine comprehension has also been explored~\cite{Ribeiro2020BeyondAB, Wan2023BiasAskerMT}.
Our paper studies the robustness of another widely-used AI software, namely multi-modal content moderation, which has not been systematically discussed.

\subsection{Testing Content Moderation Software}

Previous literature on the robustness of content moderation applications involves a diverse range of areas, including software engineering, natural language processing, computer vision and speech signal processing.
These papers have spared effort on measuring the reliability of textual content moderation software~\cite{Ahlgren2021TestingWE, Li2019TextBuggerGA, Gao2018BlackBoxGO}.
For example, Wang et al.~\cite{Wang2023MTTMMT} proposed a metamorphic testing framework for textual content moderation software and designed 11 metamorphic relations to find failure cases.
Rottger et al.~\cite{Rttger2020HateCheckFT} proposed a suite of functional tests, with 29 model functionalities, for hate speech detection models.

However, all of the above papers focused on whether the content moderation software is robust to human-intentional perturbation to text.
Our work, on the other hand, pays attention to a new paradigm of detecting the images that potentially contain toxic texts, which has not been studied before.

\subsection{Testing Optical Character Recognition System}

Another line of work that is related to this paper is the robustness of the Optical Character Recognition (OCR) system, since a straightforward idea to extract texts from images is using OCR systems.
Previous work mainly focused on adopting the adversarial attack algorithm on OCR models, either white-box~\cite{Song2018FoolingOS, Xu2020WhatMS, Yang2021CostEffectiveAA} or black-box~\cite{Bayram2022ABA}, aiming to find a small and imperceptible perturbation to add on the image so that the OCR model cannot recognize the text in the image correctly.
Besides, Chen and Xu~\cite{Chen2020AttackingOC} studied the adversarial robustness of the OCR model to watermarks.

Considering \methodname's comprehensiveness, we believe it still contributes a lot compared to the OCR testing papers.
Among our MRs, only two are similar to the perturbations in these papers.
To the best of our knowledge, the remaining 19 MRs in \methodname have not been explored yet in previous literature.
Additionally, we derive the MRs with our pilot study on real user inputs, which distinguish our \methodname from related work that leverages adversarial attack algorithms to add perturbations.
Last but not least, most existing papers only evaluated the proposed methods on academic models, while in this paper we also assess \methodname on three commercial content moderation products.
Thus, we believe \methodname is the first comprehensive testing framework for textual toxic contents spread through images.

\section{Conclusion}

This paper proposed a comprehensive testing framework \methodname for validating  content moderation software.
Unlike existing testing or adversarial attack technique for NLP software, which only provide common perturbations and cover a very limited set of toxic inputs that malicious users may produce, \methodname contains 21 metamorphic relations that are mainly inspired by a pilot study. 
In addition, all the metamorphic relations in \methodname have been implemented for two languages: English and Chinese.
Our evaluation shows that the test cases generated by \methodname can easily evade the moderation of two SOTA moderation algorithms and commercial content moderation software provided by Google, Microsoft, Baidu, Tencent, and Alibaba.
The test cases have been utilized to retrain the algorithms, which exhibited substantial improvement in model robustness while maintaining identical accuracy on the original test set.
We believe that this work is the crucial first step toward systematic testing of content moderation software.

\section{Acknowledgement}

The work described in this paper was supported by the Key-Area Research and Development Program of Guangdong Province (No. 2020B010165002) and the Key Program of Fundamental Research from Shenzhen Science and Technology Innovation Commission (No. JCYJ20200109113403826). It was also supported by the Research Grants Council of the Hong Kong Special Administrative Region, China (No. CUHK 14206921 of the General Research Fund).

\bibliographystyle{IEEEtran}
\bibliography{reference}

\end{document}